\documentclass[preprint,prd,showpacs,showkeys,byrevtex,footinbib,eqsecnum,unsortedaddress]{revtex4}
\usepackage{amsfonts}
\usepackage{amsmath}
\usepackage{amssymb}

\input{tcilatex}

\begin{document}

\title{Infrared behavior of the dispersion relations in high-temperature
scalar QED}
\author{A. Abada}
\email{a.abada@uaeu.ac.ae}
\affiliation{Physics Department, United Arab Emirates University, POB 17551, Al Ain,
United Arab Emirates}
\altaffiliation{On leave from: D\'{e}partement de Physique, ENS, BP 92 Vieux Kouba, 16050
Alger, Algeria}
\author{K. Bouakaz}
\email{bouakazk@caramail.com}
\affiliation{D\'{e}partement de Physique, Ecole Normale Sup\'{e}rieure, BP 92 Vieux
Kouba, 16050 Alger, Algeria}

\begin{abstract}
We investigate the infrared properties of the next-to-leading-order
dispersion relations in scalar quantum electrodynamics at high temperature
in the context of hard-thermal-loop perturbation theory. Specifically, we
determine the damping rate and the energy for scalars with ultrasoft
momenta. We show by explicit calculations that an early external-momentum
expansion, before the Matsubara sum is performed, gives exactly the same
result as a late one. The damping rate is obtained up to fourth order
included in the ultrasoft momentum and the energy up to second order. The
damping rate is found sensitive in the infrared whereas the energy not.
\end{abstract}

\date{\today}
\maketitle

\section{Introduction}

Because of asymptotic freedom, one thinks that the structure and properties
of high-temperature quantum chromodynamics (QCD) should be accessible by
perturbative means, however these are organized \cite{kraemmer-rebhan}.
Originally, the standard loop-expansion \cite{lan,Kap} ran into difficulties
when attempting to describe the properties of slow-moving particles: gauge
invariance was lost \cite{sil}, a consequence of the fact that the standard
loop-expansion did not reflect anymore a systematic expansion in powers of
the small coupling constant. A re-organization of lowest-order terms in the
correlation functions was therefore necessary and indeed performed: the
so-called hard thermal loops (HTL) were systematically extracted from
one-loop order diagrams and added to the lowest-order quantities \cite%
{BP1,BP2,BP3,Pi1,Pi2, Pi3,Fren-Taylor,le bellac}. It should also be
mentioned that the need for resummation of standard loop-expansion is not
peculiar to non-abelian theories but is also necessary in the Abelian case,
for quantum electrodynamics (QED) \cite{le bellac} \ and scalar QED \cite%
{kraemmer-rebhan-schulz}, as well as for the simpler $\varphi ^{4}$-theory 
\cite{parwani}. Thus are formed dressed propagators and vertices and a
convenient description of the slow-moving quasi-particles becomes possible.
Next-to-leading order quantities, in this new framework, are obtained via
one-loop diagrams involving these dressed propagators and vertices. In this
new context, some of the old difficulties are cured but others persist, most
notably sensitivity to the infrared and the light-cone, essentially due to
the fact that static (chromo)magnetic fields remain unshielded even after
the HTL reorganization. In QCD, these are thought of becoming screened at
the so-called magnetic scale, which is believed to manifest itself
non-perturbatively in the dispersion relations at next-to-leading order.
Because of this, perturbation is believed to break down at some order, which
depends on the quantity under consideration \cite%
{linde1,linde2,gross-pisarski-yaffe}.

Still, useful information can be extracted from the perturbative regime.
Indeed, using this method, an early explicit calculation of the
zero-momentum transverse gluon damping rate shows that it is finite and
positive \cite{gamt}. A similar computation for the quarks has also been
carried out independently in \cite{kobes-kunstatter-mak} and \cite{BP
(quarks)} and it yields a finite positive number too. As for the calculation
of static quantities such as free energies and screening masses, these are
most conveniently carried out in the imaginary-time formalism since analytic
continuation to real time is not necessary \cite{Blaisot-Iancu-Parwani}
anymore. For such quantities, there exists a simplified resummation scheme 
\cite{Arnold-Espinosa}, based on the fact that in Euclidean formalism, the
momentum can be soft only for the zero mode in the Matsubara sum. Hence
dressing is necessary only for the static propagators, and calculations
based on this are relatively simpler to carry out than those of the full
resummation, necessary when considering dynamic quantities. The `reduced'
approach has been used in many instances like in the study of phase
transitions in gauge theories \cite{Arnold-Espinosa}, the calculation of the
free energy in $\varphi ^{4}$-theory and QCD \cite%
{Arnold-Zhai,Zhai-Kastening} as well as in the computation of the electric
screening mass in QED and scalar QED \cite{Blaisot-Iancu-Parwani}. Other
important quantities in this regard are the next-to-leading order Debye
screening length \cite{rebhan-94-93} and next-to-leading order correction to
the gluonic plasma frequency \cite{schulz-1}.

But as mentioned, difficulties still persist after HTL reorganization. For
example, logarithmic sensitivity is encountered in an early estimation of
the damping rate for a heavy fermion \cite{Pi1}, and more generally for
fast-moving particles \cite{Leb90,Leb91,Burgess-Marini,Rebhan}. Also, an
estimation of the damping of moving quasi-particles is logarithmically
dependent on the coupling \cite{Pi-93,Flechsig-Rebhan-Schulz}. The infrared
problem was also emphasized when attempting to estimate the damping rates of
non-moving longitudinal gluons \cite{AAB,AA} and slow-moving quarks \cite%
{ABD}.

What is of concern here is, partly, the analytic properties of QCD at finite
temperature in the infrared. In order to investigate some of these
properties, the context of scalar QED is useful because it offers a simpler
setting in which calculations are facilitated by the fact that there are no
hard thermal loops in the vertices and those related to the scalar
self-energy are momentum independent \cite{kraemmer-rebhan-schulz}. This
situation allows sometimes carrying out almost complete momentum-dependent\
calculations. Of particular interest to us is an issue raised by the works 
\cite{AAB,AA}: in order to carry forward analytically, one had to expand in
powers of the external momentum $p$, considered ultrasoft to ensure the
validity of the expansion, to allow one to perform intermediary solid-angle
integrals which would have been, otherwise, intractable analytically. The
particular feature of this expansion is that it was done in the
imaginary-time formalism, before the Matsubara sum was performed and the
analytic continuation to real energies taken. The subsequent appearance of
infrared sensitivity in the final coefficients may then be linked to this
`early expansion'.

To tackle this specific issue in more depth, we embark in this work on the
investigation of the next-to-leading order dispersion relation for scalars
with ultrasoft momentum $p$ in the context of next-to-leading order HTL
perturbation of scalar QED. We will examine both the damping rate $\gamma
_{s}\left( p\right) $ and the energy $\omega _{s}\left( p\right) $. Note
that the damping rate $\gamma _{s}\left( p\right) $ has already been
investigated in \cite{thoma} and found, to second order in $p$,
logarithmically infrared sensitive. In this work, the expansion of $\gamma
_{s}\left( p\right) $ is pushed to fourth order in $p$ and that of $\omega
_{s}\left( p\right) $ to second order. What is interesting is that we
perform the expansions by two different methods: (i) We delay the momentum
expression until after the Matsubara sum is performed and the analytic
continuation to real energies taken. As for the damping rate $\gamma
_{s}\left( p\right) $, we will find the coefficient of second order
logarithmically sensitive to the infrared cut-off $\eta $ as in \cite{thoma}
whereas the coefficient of fourth order will behave like $1/\eta ^{2}$
(odd-order coefficients cancel). As for the energy $\omega _{s}\left(
p\right) $, no infrared sensitivity will appear. (ii) We perform the
early-momentum expansion and carry on with steps similar to \cite{AAB,AA}.
We find exactly the same results as those of the previous method. This must
alleviate some pressure on the early-momentum expansion method: it may not
be responsible for the appearance of infrared sensitivity. In the context of
QCD, there is simply little hope of obtaining compact analytic expressions,
and expansions such as the early-momentum one are sometimes necessary.

This article is organized as follows. After this introduction, the second
section recapitulates the hard thermal loops and the dressing of the
propagators. Section three discusses the calculation of the scalar damping
rate $\gamma _{s}\left( p\right) $ by the two methods and section four the
calculation of the scalar energy $\omega _{s}\left( p\right) $. Section five
is devoted to the discussion of the results and includes some concluding
remarks.

\section{Dressing the propagators and the dispersion relation}

Let us first recapitulate the results giving the propagators and vertices to
be used in the forthcoming next-to-leading order calculations. We use the
imaginary-time formalism in which the four-momentum is $P^{\allowbreak \mu
}=(p_{0},\mathbf{p})$ such that $P^{2}=p_{0}^{2}+p^{2}$ with\ the scalar
Matsubara frequency$\ p_{0}=2\pi nT$, $n$ an integer and $T$ the
temperature. After the evaluation of all frequency sums, $p_{0}$ is
analytically continued to the real external energy $\omega $ using the
analytic continuation $p_{0}=-i\omega +0^{+}$.

\subsection{Dressed propagators}

To leading order, the self-energies are obtained from undressed one-loop
diagrams. For soft external momenta, that is for $\omega $ and $p$ of order $%
eT$ where $e$ is the coupling constant, the dominant contributions to these
diagrams come from loop momenta of the order of $T$, the hard scale. These
contributions are called hard thermal loops. For the scalar self-energy, the
hard thermal loop is given by \cite{kraemmer-rebhan-schulz}:%
\begin{equation}
\delta \Sigma \left( P\right) =-m_{s}^{2},  \label{HTL-scalar-self-energy}
\end{equation}%
with $m_{s}=eT/2$, the scalar thermal mass. Note that $\delta \Sigma $ is
momentum independent. When dressed with the hard thermal loop, the
leading-order propagator for the scalar becomes:%
\begin{equation}
^{\ast }\Delta _{s}\left( P\right) =\frac{1}{P^{2}+m_{s}^{2}}\,.
\label{dressed scalar propagator}
\end{equation}

The hard thermal loop $\delta \Pi ^{\mu \upsilon }$ in the photon
self-energy is given in \cite{thoma} and can be expressed in terms of two
independent scalar functions $\delta \Pi _{l}(K)$ and $\delta \Pi _{t}(K)$
given by:%
\begin{equation}
\delta \Pi _{l}(K)=3\,m_{p}^{2}\,Q_{1}\left( \frac{ik_{0}}{k}\right) ;\text{
\ \ }\delta \Pi _{t}(K)=\frac{3}{5}m_{p}^{2}\left[ Q_{3}\left( \frac{ik_{0}}{%
k}\right) -Q_{1}\left( \frac{ik_{0}}{k}\right) -\frac{5}{3}\right] ,
\label{photon-self-energy}
\end{equation}%
where $m_{p}=eT/3$ is the photon thermal mass and the $Q_{i}$'s are Legendre
functions of the second kind. In the strict Coulomb gauge, the components of
the dressed photon propagator $^{\ast }\Delta _{\allowbreak \mu \nu }\left(
K\right) $ are as follows:%
\begin{align}
^{\ast }\Delta _{00}\left( K\right) & =\,^{\ast }\Delta _{l}\left( K\right) ,%
\text{ \ \ \ }^{\ast }\Delta _{0i}\left( K\right) =0;  \notag \\
^{\ast }\Delta _{ij}\left( K\right) & =\left( \delta _{ij}-\hat{k}_{i}\hat{k}%
_{j}\right) \,^{\ast }\Delta _{t}\left( K\right) ,
\label{effective-photon-propagator}
\end{align}%
where $^{\ast }\Delta _{l}$ and $^{\ast }\Delta _{t}$ are the propagators
for the longitudinal and transverse photons respectively. They have the
following expressions: 
\begin{equation}
^{\ast }\Delta _{l}\left( K\right) =\frac{1}{k^{2}-\delta \Pi _{l}\left(
K\right) }\,;\,\,\,\,\,\,\,\,\,\,\,\,\,\,\,\,\,\,\,\,^{\ast }\Delta
_{t}\left( K\right) =\frac{1}{K^{2}-\delta \Pi _{t}\left( K\right) }\,.
\label{photon propagators}
\end{equation}

Now one peculiarity of scalar QED is that the vertices remain undressed,
i.e., unaffected by the hard thermal loops \cite{kraemmer-rebhan-schulz}.
The vertex with one photon and two scalar external lines ($Q$ incoming, $P$
outgoing) is:%
\begin{equation}
\Gamma ^{\allowbreak \mu }\left( P,Q\right) =-e\left( P+Q\right)
^{\allowbreak \mu }\,,  \label{eff-ver-1}
\end{equation}%
and the vertex between two photons and two scalars is:%
\begin{equation}
\Gamma ^{\allowbreak \mu \nu }\left( P,Q\right) =2e^{2}\,\delta
^{\allowbreak \mu \nu }.  \label{eff-ver-2}
\end{equation}

\subsection{Dispersion relation}

The scalar damping rate $\gamma _{s}\left( p\right) $ and energy $\omega
_{s}\left( p\right) $ are obtained from the scalar complex energy $\omega $
which satisfies the following full scalar dispersion relation \cite{thoma}:%
\begin{equation}
\omega ^{2}=p^{2}-\Sigma \left( \omega ,p\right) ,
\label{full-dispersion-in-SQED}
\end{equation}%
where $\Sigma \left( \omega ,p\right) $ is the full scalar self-energy.
Taking account of the next-to-leading order terms in $e$ and expanding
everything around the leading order\ result $\omega _{s0}\left( p\right) $
(squared) for fixed $p$, we obtain:%
\begin{equation}
\omega ^{2}=p^{2}-\delta \Sigma \left( \omega _{s0},p\right) -\,^{\ast
}\Sigma \left( \omega _{s0},p\right) -\left( \omega ^{2}-\omega
_{s0}^{2}\right) \left. \partial _{x^{2}}\delta \Sigma \left( x,p\right)
\right\vert _{x=\omega _{s0}}+\mathcal{O}\left( e^{4}T^{2}\right) .
\label{next-to-leading-dispersion}
\end{equation}%
In this relation, $\omega _{s0}\left( p\right) =\sqrt{m_{s}^{2}+p^{2}}$,
pole of $^{\ast }\Delta _{s}\left( P\right) $ from (\ref{dressed scalar
propagator}) and solution to (\ref{full-dispersion-in-SQED}) to lowest
order. $\delta \Sigma $ is nothing but the hard thermal loop given in (\ref%
{HTL-scalar-self-energy}) and $^{\ast }\Sigma $ is the next-to-leading order
contribution. Using (\ref{HTL-scalar-self-energy}), we can rewrite (\ref%
{next-to-leading-dispersion}) as:%
\begin{equation}
\omega ^{2}=\omega _{s0}^{2}-~^{\ast }\Sigma \left( \omega _{s0},p\right) +%
\mathcal{O}\left( e^{4}T^{2}\right) .
\label{next-to-leading-dispersion-rewritten}
\end{equation}

The full energy $\omega \left( p\right) $ is in general complex. If we
denote its real part (the scalar energy) by $\omega _{s}\left( p\right) $,
then we have:%
\begin{equation}
\omega _{s}\left( p\right) =\omega _{s0}-\frac{\func{Re}\delta \hspace{1pt}%
^{\ast }\Sigma \left( \omega _{s0},p\right) }{2\omega _{s0}}+\mathcal{O}%
\left( e^{3}T\right) .  \label{next-to-leading-order-scalar-energy-1}
\end{equation}%
The prefix $\delta $ before $^{\ast }\Sigma $ in (\ref%
{next-to-leading-order-scalar-energy-1}) indicates that the known leading
hard-thermal-loop contribution has to be subtracted for ultraviolet
convergence \cite{kraemmer-rebhan-schulz}. The damping rate for scalars is
defined by $\gamma _{s}(p)=-\mathrm{\func{Im}}\omega \left( p\right) $.\ It
is $e$-times smaller than $\omega _{s0}(p)$, and so we have to lowest order 
\cite{thoma}:%
\begin{equation}
\gamma _{s}\left( p\right) =\frac{1}{2\omega _{s0}}\func{Im}\,^{\ast }\Sigma
\left( \omega _{s0},p\right) +\mathcal{O}\left( e^{3}T\right) .
\label{gamma_s}
\end{equation}%
We see that determining $\gamma _{s}\left( p\right) $ to lowest order in $e$
and $\omega _{s}\left( p\right) $ to next-to-leading order amounts to
calculating the imaginary and real parts of the next-to-leading order scalar
self-energy $^{\ast }\Sigma $. From now on, we assume the scalar momentum\ $%
p $ \textit{ultrasoft}, i.e., of the order of $e^{2}T$. Also, we will take $%
m_{s}\equiv 1$ in the sequel, to ease the notation. We will reintroduce it
back in the final expressions.

\section{Scalar damping rate}

We start by calculating the damping rate $\gamma _{s}\left( p\right) $. The
self-energy $^{\ast }\Sigma \left( P\right) $ is the sum of two
contributions \cite{thoma}:%
\begin{equation}
^{\ast }\Sigma \left( P\right) =\text{ }^{\ast }\Sigma _{1}\left( P\right)
+~^{\ast }\Sigma _{2}\left( P\right) ,  \label{sigma_1+2}
\end{equation}%
where the contribution $^{\ast }\Sigma _{1}$ is a one-loop dressed diagram
with two\ one-photon-two-scalar vertices given in (\ref{eff-ver-1}):%
\begin{equation}
^{\ast }\Sigma _{1}\left( P\right) =\mathrm{Tr}_{\mathrm{soft}}\left[ \Gamma
^{\mu }\left( P,Q\right) \,^{\ast }\Delta _{s}\left( Q\right) \Gamma ^{\nu
}\left( Q,P\right) \,^{\ast }\Delta _{\allowbreak \mu \nu }\left( K\right) %
\right] ,  \label{eff-self-energy-1}
\end{equation}%
and the contribution $^{\ast }\Sigma _{2}$ is the dressed tadpole diagram
with a two-photon-two-scalar vertex given in (\ref{eff-ver-2}): 
\begin{equation}
^{\ast }\Sigma _{2}\left( P\right) =\mathrm{Tr}_{\mathrm{soft}}\left[ \Gamma
^{\mu \nu }\left( K,P\right) \,^{\ast }\Delta _{\allowbreak \mu \nu }\left(
K\right) \right] .  \label{eff-self-energy-2}
\end{equation}%
In the above two relations, $K$ is the internal photon loop-momentum, $Q=P-K$
and $\mathrm{Tr}\equiv T\sum_{k_{0}}\int d^{3}k$ /$(2\pi )^{3}$ with $%
k_{0}=2n\pi T$. The subscript `soft' indicates that only soft momenta are
allowed in the integral. To take account of potential infrared sensitivity,
we will introduce an infrared cutoff $\eta $ in the $k$-integration.

It turns out that $^{\ast }\Sigma _{2}\left( P\right) $ is real. This is
because there is only one dressed propagator involved in its expression (\ref%
{eff-self-energy-2}) and the vertex there is undressed. It will also turn
out to be $p$-independent, see (\ref{sigma_2 final}). Therefore, $^{\ast
}\Sigma _{2}\left( P\right) $ does not contribute to the damping rate $%
\gamma _{s}\left( p\right) $; it will only shift the mass $m_{s}$ of the
scalar. We thus can simply write:%
\begin{equation}
\gamma _{s}\left( p\right) =\frac{1}{2\omega _{s0}}\func{Im}\,^{\ast }\Sigma
_{1}\left( \omega _{s0},p\right) +\mathcal{O}\left( e^{3}T\right) ,
\label{gamma_s from sigma1}
\end{equation}%
and henceforth, we will not manipulate $^{\ast }\Sigma _{2}\left( P\right) $
until we reach the calculation of the scalar energy $\omega _{s}\left(
p\right) $ in section four.

\subsection{Late momentum expansion}

In this subsection, we will first perform the Matsubara sums and
analytically continue to real energies before expanding in powers of $p$.
Using the structure (\ref{effective-photon-propagator}) of the photon
propagator in the strict Coulomb gauge and the expression of the
one-photon-two-scalar vertex (\ref{eff-ver-1}), we see that $^{\ast }\Sigma
_{1}(P)$ is composed of two terms. The first term involves longitudinal
photons and is denoted consequently by $^{\ast }\Sigma _{1l}(P)$, and the
second one involves transverse photons and is denoted by $^{\ast }\Sigma
_{1t}(P)$. Let us look first at $^{\ast }\Sigma _{1l}(P)$. We have:%
\begin{equation}
^{\ast }\Sigma _{1l}(P)=e^{2}T\sum_{k_{0}}\int \frac{d^{3}k}{\left( 2\pi
\right) ^{3}}\left[ \left( 2p_{0}-k_{0}\right) ^{2}\,^{\ast }\Delta
_{l}\left( K\right) \,^{\ast }\Delta _{s}\left( Q\right) \right] .
\label{sigma-star-l-first-expression}
\end{equation}%
Writing explicitly the integral over the solid angle of $\mathbf{\hat{k}}$
in a reference frame where $\mathbf{\hat{p}}$ is the principal axis, we have:%
\begin{equation}
^{\ast }\Sigma _{1l}(P)=\frac{e^{2}}{4\pi ^{2}}\,T\sum_{k_{0}}\int_{%
\allowbreak \eta }^{+\infty }dk\,k^{2}\int_{-1}^{+1}dx\left[ \left(
2p_{0}-k_{0}\right) ^{2}\,^{\ast }\Delta _{l}\left( k_{0},k\right) \,^{\ast
}\Delta _{s}\left( q_{0},q\right) \right] ,  \label{sigma-star-l}
\end{equation}%
with $x=\mathbf{\hat{k}.\hat{p}}$. Note that $q=\sqrt{k^{2}-2pkx+p^{2}}$,
which makes the integral over $x$ not feasible for the moment.

We want to perform the Matsubara sum over $k_{0}$. For this, we use the
spectral decomposition of the two dressed propagators. In general, we have 
\cite{Pi (rho)}:%
\begin{equation}
\Delta _{i}(k_{0},k)=\int_{0}^{1/T}d\tau \,e^{ik_{0}\tau }\int_{-\infty
}^{+\infty }d\omega \,\rho _{i}(\omega ,k)\left( 1+n(\omega )\right)
e^{-\omega \tau },  \label{spectral}
\end{equation}%
where $i$ stands for $l,t$ or $s$ and $n(\omega )$ is the Bose-Einstein
distribution. The explicit expressions of the spectral densities $\rho _{i}$
are displayed in (\ref{rho_l-t})-(\ref{rho_s}) below. Making this
replacement, the sum over$\ k_{0}$ can now be performed. The subsequent
steps are standard: One imaginary-time integration is eliminated by a delta
function and the second one yields an energy denominator. Everywhere except
in the energy denominator $p_{0}$ is replaced by $2\pi nT$. The analytic
continuation to real energies is taken at this stage and obtained by the
replacement $ip_{0}\rightarrow \omega _{s0}(p)+i0^{+}$. The imaginary part
is extracted using the known relation $1/\left( x+i0^{+}\right) =\Pr \left(
1/x\right) -i\pi \delta (x)$. We thus have: 
\begin{equation}
\func{Im}\,^{\ast }\Sigma _{1l}(P)=-\frac{e^{2}T}{4\pi }\hspace{-3pt}%
\int_{\allowbreak \eta }^{+\infty }\hspace{-3pt}\hspace{-3pt}\hspace{-3pt}%
\hspace{-3pt}dk\,k^{2}\hspace{-3pt}\hspace{-3pt}\int_{-1}^{+1}\hspace{-3pt}%
\hspace{-3pt}dx\hspace{-3pt}\hspace{-3pt}\int_{-\infty }^{+\infty }\hspace{%
-3pt}\hspace{-3pt}d\omega \,\frac{\omega _{s0}\left( 2\omega _{s0}-\omega
\right) ^{2}}{\omega \left( \omega _{s0}-\omega \right) }\,\,\rho _{l}\left(
\omega ,k\right) \rho _{s}\left( \omega _{s0}-\omega ,q\right) .
\label{sigma-star-l-compact}
\end{equation}%
In the above relation, only soft values of $\omega $ are to contribute in
the integral, and so we have used the approximation $n\left( \omega \right)
\simeq T/\omega $.

The transverse contribution $^{\ast }\Sigma _{1t}$ is handled in similar
steps and we obtain for its imaginary part the following expression: 
\begin{equation}
\func{Im}\,^{\ast }\Sigma _{1t}(P)=\frac{e^{2}T}{\pi }p^{2}\hspace{-3pt}%
\int_{\allowbreak \eta }^{+\infty }\hspace{-3pt}\hspace{-3pt}\hspace{-3pt}%
\hspace{-3pt}dk\,k^{2}\hspace{-3pt}\hspace{-3pt}\int_{-1}^{+1}\hspace{-3pt}%
\hspace{-3pt}dx\left( 1\hspace{-3pt}-x^{2}\right) \hspace{-3pt}\hspace{-3pt}%
\int_{-\infty }^{+\infty }\hspace{-3pt}\hspace{-3pt}d\omega \,\frac{\omega
_{s0}}{\omega \left( \omega _{s0}\hspace{-3pt}-\omega \right) }\,\rho
_{t}\left( \omega ,k\right) \rho _{s}\hspace{-3pt}\left( \omega _{s0}-%
\hspace{-3pt}\omega ,q\right) .  \label{sigma-star-t-compact}
\end{equation}%
Note that the transverse contribution to $\gamma _{s}\left( p\right) $
already starts at order $p^{2}$.

Now dividing by $2\omega _{s0}\left( p\right) $ as required in (\ref{gamma_s
from sigma1}), we get the following longitudinal and transverse
contributions to the damping rate:%
\begin{eqnarray}
\gamma _{sl}\left( p\right)  &=&-\frac{e^{2}T}{8\pi }\int_{\allowbreak \eta
}^{+\infty }dk\,k^{2}\int_{-1}^{+1}dx\,\int_{-\infty }^{+\infty }d\omega 
\frac{\left( 2\omega _{s0}-\omega \right) ^{2}\,}{\omega \left( \omega
_{s0}-\omega \right) }\,\rho _{l}\left( \omega ,k\right) \rho _{s}\left(
\omega _{s0}-\omega ,q\right) ;  \notag \\
\gamma _{st}\left( p\right)  &=&\frac{e^{2}T}{2\pi }p^{2}\int_{\allowbreak
\eta }^{+\infty }\hspace{-3pt}\hspace{-3pt}\hspace{-3pt}\hspace{-3pt}%
dk\,k^{2}\hspace{-3pt}\int_{-1}^{+1}\hspace{-3pt}dx\,\int_{-\infty
}^{+\infty }\hspace{-3pt}\hspace{-3pt}\hspace{-3pt}d\omega \frac{\left(
1-x^{2}\right) }{\omega \left( \omega _{s0}-\omega \right) }\,\rho
_{t}\left( \omega ,k\right) \rho _{s}\left( \omega _{s0}-\omega ,q\right) .
\label{scalar-damping-rate-compact-expression}
\end{eqnarray}%
According to (\ref{gamma_s from sigma1}), the damping rate itself will be:%
\begin{equation}
\gamma _{s}\left( p\right) =\gamma _{sl}\left( p\right) +\gamma _{st}\left(
p\right) .  \label{gamma_s equal gamma_l+gamma_t}
\end{equation}

Next we move to perform the integrals involved in the expressions of (\ref%
{scalar-damping-rate-compact-expression}) above. The spectral densities $%
\rho _{l,t}$ are known \cite{Pi (rho),le bellac}: 
\begin{equation}
\rho _{l,t}\left( \omega ,k\right) =\mathfrak{z}_{l,t}\left( k\right) \left[
\delta \left( \omega -\omega _{l,t}\left( k\right) \right) -\delta \left(
\omega +\omega _{l,t}\left( k\right) \right) \right] +\beta _{l,t}\left(
\omega ,k\right) \,\Theta \left( k^{2}-\omega ^{2}\right) ,  \label{rho_l-t}
\end{equation}%
where $\mathfrak{z}_{l,t}\left( k\right) $ are the residue functions and $%
\beta _{l,t}\left( \omega ,k\right) $ the cut functions. The residue
functions are given by:%
\begin{equation}
\mathfrak{z}_{l}(k)=\left. -\frac{\omega \left( \omega ^{2}-k^{2}\right) }{%
k^{2}\left( 4/3-\omega ^{2}+k^{2}\right) }\right\vert _{\omega =\omega
_{l}(k)};\qquad \mathfrak{z}_{t}(k)=\left. \frac{\omega \left( \omega
^{2}-k^{2}\right) }{4\omega ^{2}/3-\left( \omega ^{2}-k^{2}\right) ^{2}}%
\right\vert _{\omega =\omega _{t}(k)},  \label{z_l-t}
\end{equation}%
and the cut functions have the following expressions:%
\begin{align}
\beta _{l}(k,\omega )& =-\frac{2\omega }{3k}\left[ \left( 4/3+k^{2}-\frac{%
2\omega }{3k}\ln \frac{k+\omega }{k-\omega }\right) ^{2}+\dfrac{4\pi
^{2}\omega ^{2}}{9k^{2}}\right] ^{-1},  \notag \\
\beta _{t}(k,\omega )& =\frac{\omega \left( k^{2}-\omega ^{2}\right) }{3k^{3}%
}\left[ \left( k^{2}-\omega ^{2}+\dfrac{2\omega ^{2}}{3k^{2}}\left( 1+\dfrac{%
k^{2}-\omega ^{2}}{2k\omega }\ln \dfrac{k+\omega }{k-\omega }\right) \right)
^{2}\hspace{-2pt}+\left( \dfrac{\pi \omega \left( k^{2}-\omega ^{2}\right) }{%
3k^{3}}\right) ^{2}\right] ^{-1}\hspace{-2pt}.  \label{beta_l-t}
\end{align}%
Remember that all quantities are in units of $m_{s}$. The spectral density $%
\rho _{s}$ does not have a cut. It\ simply writes: 
\begin{equation}
\rho _{s}\left( \omega ,k\right) =\mathfrak{z}_{s}\left( k\right) \left[
\delta \left( \omega -\omega _{s0}\left( k\right) \right) -\delta \left(
\omega +\omega _{s0}\left( k\right) \right) \right] ,  \label{rho_s}
\end{equation}%
with $\mathfrak{z}_{s}\left( k\right) =1/2\omega _{s0}\left( k\right) $.

We therefore replace the spectral densities by their respective expressions (%
\ref{rho_l-t}) and (\ref{rho_s}). The integration over $\omega $ disappears
with the delta functions of (\ref{rho_s}). From the expression of $\rho
_{l,t}$ in (\ref{rho_l-t}), there are going to be two kinds of
contributions, i.e., a $\delta $-contribution and a $\Theta $-contribution.
Because of kinematics, the $\delta $-contribution is always zero whereas the 
$\Theta $-contribution survives. The integrations over $x$ and $k$ have to
be performed numerically. One would fit the behaviors of $\gamma _{sl}\left(
p\right) $ and $\gamma _{st}\left( p\right) $ to several ultrasoft values of 
$p$, but in the spirit of the present work and in order to make direct
comparison with the\ early momentum expansion method to be presented
shortly, we display the results for $\gamma _{sl}\left( p\right) $ and $%
\gamma _{st}\left( p\right) $ in powers of $p$ to fourth order. We obtain:%
\begin{eqnarray}
\gamma _{sl}\left( p\right)  &=&\frac{e^{2}T}{16\pi }\left[
1.44253+0.309278\,\bar{p}^{2}-0.133949\bar{p}^{4}+\mathcal{O}\left( \bar{p}%
^{6}\right) \right] +\mathcal{O}\left( e^{3}T\right) ;  \notag \\
\gamma _{st}\left( p\right)  &=&\left. \hspace{-0.2cm}\frac{e^{2}T}{4\pi }%
\right[ -\left( 1.65937+1.62114\ln \allowbreak \bar{\eta}\right) \bar{p}^{2}
\notag \\
&&+\left. \left( 71.3264+57.4152\ln \allowbreak \bar{\eta}+\frac{1.94537}{%
\allowbreak \bar{\eta}^{2}}\right) \bar{p}^{4}+\mathcal{O}\left( \bar{p}%
^{6}\right) \right] +\mathcal{O}\left( e^{3}T\right) .  \label{gamma-result}
\end{eqnarray}%
The thermal mass $m_{s}$ has been reintroduced and here, $\bar{p}=p/m_{s}$
and $\allowbreak \bar{\eta}=\eta /m_{s}$.

To order $\bar{p}^{2}$, these are the results obtained already in \cite%
{thoma}. Note that the longitudinal contribution $\gamma _{sl}\left(
p\right) $ is safe from any infrared sensitivity whereas the transverse
contribution $\gamma _{st}\left( p\right) $ is infrared sensitive, this
despite the fact that both contributions have been handled in exactly
similar ways. The reason why we have wanted to push the expansion to order $%
\bar{p}^{4}$ is to show that other forms of infrared sensitivity may appear,
other than the familiar $\ln \bar{\eta}$. Indeed, we clearly see the
power-like behavior $1/\bar{\eta}^{2}$ in the $\bar{p}^{2}$-coefficient of $%
\gamma _{st}\left( p\right) $. These issues will be furthered in section
five. According to (\ref{gamma_s equal gamma_l+gamma_t}), the scalar damping
rate is:%
\begin{eqnarray}
\gamma _{s}\left( p\right)  &=&\left. \frac{e^{2}T}{16\pi }\right[ \hspace{%
-0.1cm}1.44253-\left( 6.32820+6.48456\ln \allowbreak \bar{\eta}\right) \bar{p%
}^{2}  \notag \\
&&+\left. \left( 285.172+229.661\ln \allowbreak \bar{\eta}+\frac{7.78148}{%
\allowbreak \bar{\eta}^{2}}\right) \bar{p}^{4}+\mathcal{O}\left( \bar{p}%
^{6}\right) \right] +\mathcal{O}\left( e^{3}T\right) .  \label{gamma_s final}
\end{eqnarray}

\subsection{Early momentum expansion}

Now let us perform the same calculation while introducing from the start an
early momentum expansion, before the Matsubara sum and analytic continuation
to real energies are done. First we perform analytically the integrals over
the solid angle of $\mathbf{\hat{k}}$ in the same reference frame where $%
\mathbf{\hat{p}}$ is the principal axis. The difficulty comes from the
presence of functions of $q=\sqrt{k^{2}-2pkx+p^{2}}$, mainly the scalar
dressed propagator $^{\ast }\Delta _{s}\left( q_{0},q\right) $. These need
to be expanded and, in order to do this, we use the following expansion to
fourth order in the external momentum $p$:%
\begin{eqnarray}
^{\ast }\Delta _{s}\left( q_{0},q\right)  &=&\left[ 1-px\,\partial _{k}+%
\frac{p^{2}}{2}\left[ \frac{1-x^{2}}{k}\partial _{k}+x^{2}\partial _{k}^{2}%
\right] \right.   \notag \\
&&\hspace{-36pt}+\frac{p^{3}x}{2}\left[ \frac{1-x^{2}}{k^{2}}\partial _{k}-%
\frac{1-x^{2}}{k}\partial _{k}^{2}-\frac{x^{3}}{3}\partial _{k}^{3}\right] +%
\frac{p^{4}}{4k}\left[ \frac{-\left( 1\hspace{-2pt}-6x^{2}\hspace{-2pt}%
+5x^{4}\right) }{2k^{2}}\partial _{k}\right.   \notag \\
&&\hspace{-36pt}+\left. \left. \frac{\left( 1\hspace{-2pt}-6x^{2}\hspace{-2pt%
}+5x^{4}\right) }{2k}\,\partial _{k}^{2}+2x^{2}\left( 1\hspace{-2pt}%
-x^{2}\right) \partial _{k}^{3}+\frac{x^{4}\,k}{6}\partial _{k}^{4}\right] +%
\mathcal{O}\left( p^{5}\right) \right] \hspace{-2pt}\hspace{-2pt}\,^{\ast
}\Delta _{s}\left( q_{0},k\right) .  \label{expansion-delta}
\end{eqnarray}%
Here $\partial _{k}$ stands for $\partial /\partial k$ and, remember, $x=%
\mathbf{\hat{k}.\hat{p}}$. Consider first $^{\ast }\Sigma _{1l}(P)$. Insert
the above expansion in its expression (\ref{sigma-star-l-first-expression})
and then perform the integrations over $x$ which become straightforward. All
odd orders in $p$ cancel out and we are left with:%
\begin{eqnarray}
^{\ast }\Sigma _{1l}(P) &=&\frac{e^{2}}{2\pi ^{2}}T\sum_{k_{0}}\int_{%
\allowbreak \eta }^{+\infty }dk\,k^{2}\left( 2p_{0}-k_{0}\right)
^{2}\,\,^{\ast }\Delta _{l}\left( k_{0},k\right)   \notag \\
&&\times \left[ 1+\frac{p^{2}}{3}\left( \frac{1}{k}\partial _{k}+\frac{1}{2}%
\partial _{k}^{2}\right) +\frac{p^{4}}{30}\left( \partial _{k}^{3}+\frac{k}{4%
}\partial _{k}^{4}\right) +\mathcal{O}\left( p^{6}\right) \right] \,^{\ast
}\Delta _{s}\left( q_{0},k\right) .  \label{sigma_1l-expanded-1}
\end{eqnarray}%
Only at this stage the Matsubara sum and analytic continuation to real
energies are done. Here too we replace the dressed propagators by their
spectral decompositions given in (\ref{spectral}) and follow the standard
steps sketched right after (\ref{spectral}) in order to extract the
imaginary part. Dividing by $2\omega _{s0}\left( p\right) $ as indicated in (%
\ref{gamma_s}), we obtain the following expression:%
\begin{align}
\gamma _{sl}(p)& =-\frac{e^{2}T}{4\pi }\int_{\allowbreak \eta }^{+\infty }%
\hspace{-0.5cm}dk\,k^{2}\int_{-\infty }^{+\infty }d\omega \,\frac{\left(
2\omega _{s0}-\omega \right) ^{2}}{\omega \left( \omega _{s0}-\omega \right) 
}\,\rho _{l}(\omega ,k)  \notag \\
& \times \left[ 1+\frac{p^{2}}{3}\left( \frac{1}{k}\partial _{k}+\frac{1}{2}%
\partial _{k}^{2}\right) +\frac{p^{4}}{30}\left( \partial _{k}^{3}+\frac{k}{4%
}\partial _{k}^{4}\right) +\mathcal{O}\left( p^{6}\right) \right] \rho
_{s}(\omega _{s0}-\omega ,k).  \label{scalar-damping-rate-longi}
\end{align}

The transverse-photon contribution $^{\ast }\Sigma _{1t}(P)$ is handled in
similar steps. Dividing by $2\omega _{s0}\left( p\right) $, we obtain from
it the following transverse contribution to the damping rate: 
\begin{eqnarray}
\gamma _{st}(p) &=&\frac{2e^{2}T}{3\pi }p^{2}\int_{\allowbreak \eta
}^{+\infty }\hspace{-0.5cm}dk\,k^{2}\int_{-\infty }^{+\infty }\frac{d\omega 
}{\omega \left( \omega _{s0}-\omega \right) }\rho _{t}(\omega ,k)  \notag \\
&&\times \left[ 1+\frac{2p^{2}}{5}\left( \frac{1}{k}\partial _{k}+\frac{1}{4}%
\partial _{k}^{2}\right) +\mathcal{O}\left( p^{6}\right) \right] \rho
_{s}(\omega _{s0}-\omega ,k).  \label{scalar-damping-rate-trans}
\end{eqnarray}%
As one sees, there are different types of terms involved in (\ref%
{scalar-damping-rate-longi}) and (\ref{scalar-damping-rate-trans}). In the
sequel, we will briefly show how we carry through with each one. We will try
to have the notation as clear and concise as possible.

\subsubsection{Integration}

The first type of integrals and the simplest we have to deal with is the one
that involves a $\rho \rho $ contribution with no derivatives. Generically,
we consider an integral of the type: 
\begin{equation}
I_{\rho \rho }=\int_{\allowbreak \eta }^{+\infty }\hspace{-2pt}%
dk\int_{-\infty }^{+\infty }\hspace{-2pt}d\omega f(k,\omega ,1-\omega )\rho
_{i}(\omega ,k)\rho _{s}(1-\omega ,k),  \label{integral in rho rho}
\end{equation}%
where $i$ stands for $l$ (longitudinal) or $t$ (transverse). According to
the form of the spectral functions (\ref{rho_l-t}) and (\ref{rho_s}), there
are two kinds of contributions: a $\delta \delta $ (residue-residue)
contribution and a $\Theta \delta $ (residue-cut) contribution. The $\delta
\delta $ contribution requires that the energies satisfy $\pm \omega
_{s0}(k)\pm \omega _{i}(k)=1$, four constraints which are always forbidden
by the dispersion relations. Hence the $\delta \delta $ contribution is
always zero because of kinematics. The $\Theta \delta $ contribution writes $%
\int_{\allowbreak \eta }^{+\infty }\hspace{-2pt}dk\int_{-\infty }^{+\infty }%
\hspace{-2pt}d\omega f(k,\omega ,1-\omega )\,\mathfrak{z}_{s}(k)\beta
_{i}\left( k,\omega \right) \left[ \delta \left( 1-\omega -\omega
_{s0}\right) -\delta \left( 1-\omega +\omega _{s0}\right) \right] \Theta 
\hspace{-1pt}\left( k-\left\vert \omega \right\vert \right) $. A non-zero
contribution must satisfy $\omega =1\pm \omega _{s0}$, together with $-k\leq
\omega \leq k$. It is not difficult to see that only the case $\omega
=1-\omega _{s0}(k)$ is allowed, and this for all values $k\geq \allowbreak
\eta $. The integration over $\omega $ is straightforward and we obtain for
this type of integral:%
\begin{equation}
I_{\rho \rho }=\int_{\allowbreak \eta }^{+\infty }\hspace{-2pt}dk\,\mathfrak{%
z}_{s}f(k,1-\omega _{s0},\omega _{s0})\beta _{i}\left( k,1-\omega
_{s0}\right) .  \label{integral in rho- rho-result}
\end{equation}%
Note that only integrals of the type $I_{\rho \rho }$ contribute to the
coefficient of lowest order in $p$ in the scalar damping rate.

The second type of integrals we have to deal with is the following: 
\begin{equation}
I_{\rho \partial _{k}\rho }=\int_{\allowbreak \eta }^{+\infty }\hspace{-2pt}%
dk\int_{-\infty }^{+\infty }\hspace{-2pt}d\omega f(k,\omega ,1-\omega )\rho
_{i}(\omega ,k)\partial _{k}\rho _{s}(1-\omega ,k).
\label{integral in rho d_k rho}
\end{equation}%
Here too the discussion has to be carried out contribution by contribution,
using the structure of the spectral functions (\ref{rho_l-t}) and (\ref%
{rho_s}). The first contribution to consider is the one that involves two
delta functions $\int_{\allowbreak \eta }^{+\infty }\hspace{-2pt}%
dk\int_{-\infty }^{+\infty }\hspace{-2pt}d\omega \,\mathfrak{z}%
_{i}\,f(k,\omega ,1-\omega )\,\delta \left( \omega \mp \omega _{i}\right)
\partial _{k}\left[ \mathfrak{z}_{s}\delta \left( 1-\omega \mp \omega
_{s0}\right) \right] $. When the derivative over $k$ is applied to $%
\mathfrak{z}_{s}(k)$, we have zero contribution. Indeed, in order for it to
be nonvanishing, the energies must here too satisfy $\pm \omega _{s0}(k)\pm
\omega _{i}(k)=1$, forbidden by kinematics as already mentioned. But when
applying the $k$-derivative to the delta function, we also have zero
contribution. This is simply because there is no intersection between the
supports of the two delta functions, even if they are involved through
derivatives, first order as in here or higher. The other contribution is a
cut-residue, namely $\int_{\allowbreak \eta }^{+\infty }\hspace{-2pt}%
dk\int_{-\infty }^{+\infty }\hspace{-2pt}d\omega f(k,\omega ,1-\omega )\beta
_{i}\left( k,\omega \right) \Theta \hspace{-1pt}\left( k-\left\vert \omega
\right\vert \right) \partial _{k}\left[ \mathfrak{z}_{s}\delta \left(
1-\omega \mp \omega _{s0}\right) \right] $. We first apply the derivative
over $k$ to $\mathfrak{z}_{s}$ and we get the piece $\int_{\allowbreak \eta
}^{+\infty }\hspace{-2pt}dk\,\mathfrak{z}_{s}^{\prime }f(k,1-\omega
_{s0},\omega _{s0})\,\beta _{i}\left( k,1-\omega _{s0}\right) $, where $%
\mathfrak{z}_{s}^{\prime }$ stands for $d\mathfrak{z}_{s}(k)/dk$. We then
apply it to $\delta \left( 1-\omega \mp \omega _{s0}\right) $ and use the
standard rules regulating the handling of the delta distribution and we
always check the results by regularizing either the derivative $\partial
_{k} $ or the delta function itself. Only $\omega =1-\omega _{s0}$
contributes and we obtain $\int_{\allowbreak \eta }^{+\infty }\hspace{-2pt}%
dk\,\omega _{s0}^{\prime }\,\mathfrak{z}_{s}\left. \partial _{\omega }\left[
f(k,1-\omega ,\omega )\,\beta _{i}\left( k,1-\omega \right) \right]
\right\vert _{\omega =\omega _{s0}}$. Putting the above two contributions
together, we obtain the following result: 
\begin{align}
I_{\rho \partial _{k}\rho }& =\int_{\allowbreak \eta }^{+\infty }\hspace{-2pt%
}dk\left[ \mathfrak{z}_{s}^{\prime }f(k,1-\omega _{s0},\omega
_{s0})\,\,\beta _{i}\left( k,1-\omega _{s0}\right) \right.  \notag \\
& \hspace{0.6in}\left. +\omega _{s0}^{\prime }\,\mathfrak{z}_{s}\left.
\partial _{\omega }\left[ f(k,1-\omega ,\omega )\,\beta _{i}\left(
k,1-\omega \right) \right] \right\vert _{\omega =\omega _{s0}}\right] .
\label{integral in rho d_k rho-result}
\end{align}

The third type of integrals we have to deal with is one that involves a
second derivative in $k$: 
\begin{equation}
I_{\rho \partial _{k}^{2}\rho }=\int_{\allowbreak \eta }^{+\infty }\hspace{%
-2pt}dk\int_{-\infty }^{+\infty }\hspace{-2pt}d\omega f(k,\omega ,1-\omega
)\rho _{i}(k,\omega )\,\partial _{k}^{2}\rho _{s}(1-\omega ,k).
\label{integral in rho d_k2 rho}
\end{equation}%
The steps to treat the different contributions parallel those followed for $%
I_{\rho \partial _{k}\rho }$. As explained before, the $\delta \delta $
contribution is zero because of kinematics. Therefore, the only contribution
to look at is $\int_{\allowbreak \eta }^{+\infty }\hspace{-2pt}%
dk\int_{-\infty }^{+\infty }\hspace{-2pt}d\omega f(k,\omega ,1-\omega
)\,\beta _{i}\left( k,\omega \right) \Theta \hspace{-1pt}\left( k-\left\vert
\omega \right\vert \right) \partial _{k}^{2}\left[ \mathfrak{z}_{s}\,\delta
\left( 1-\omega \mp \omega _{s0}\right) \right] $. We have to use the
identity $\partial _{k}^{2}\left( \mathfrak{z}_{s}\,\delta \right) =%
\mathfrak{z}_{s}^{\prime \prime }\,\delta +2\mathfrak{z}_{s}^{\prime
}\,\partial _{k}\delta +\mathfrak{z}_{s}\,\partial _{k}^{2}\delta $, where $%
\mathfrak{z}_{s}^{\prime \prime }$ stands for the second derivative of $%
\mathfrak{z}_{s}\left( k\right) $. The term involving $\mathfrak{z}%
_{s}^{\prime \prime }\,\delta $ yields $\int_{\allowbreak \eta }^{+\infty }%
\hspace{-2pt}dk\mathfrak{z}_{s}^{\prime \prime }f(k,1-\omega _{s0},\omega
_{s0})\,\beta _{i}\left( k,1-\omega _{s0}\right) $. The two other terms $%
\mathfrak{z}_{s}^{\prime }\,\partial _{k}\delta +\mathfrak{z}_{s}\,\partial
_{k}^{2}\delta $ give together the following result $\int_{\allowbreak \eta
}^{+\infty }\hspace{-2pt}dk\left. \left[ \left( \mathfrak{z}_{s}\omega
_{s0}^{\prime \prime }+2\mathfrak{z}_{s}^{\prime }\omega _{s0}^{\prime
}\right) \partial _{\omega }+\mathfrak{z}_{s}\omega _{s0}^{\prime 2}\partial
_{\omega }^{2}\right] \left[ f(k,1-\omega ,\omega )\,\beta _{i}\left(
k,1-\omega \right) \right] \right\vert _{\omega =\omega s0}$. The procedure
is to replace $\partial _{k}$ by $\omega _{s0}\partial _{\omega }$ and apply
the usual rules regulating the handling of the delta distribution. Putting
the above results together, we obtain: 
\begin{align}
I_{\rho \partial _{k}^{2}\rho }& =\int_{\allowbreak \eta }^{+\infty }\hspace{%
-2pt}dk\left[ \mathfrak{z}_{s}^{\prime \prime }f(k,1-\omega _{s0},\omega
_{s0})\,\beta _{i}\left( k,1-\omega _{s0}\right) \right.  \notag \\
& +\left. \left[ \left( \mathfrak{z}_{s}\omega _{s0}^{\prime \prime }+2%
\mathfrak{z}_{s}^{\prime }\omega _{s0}^{\prime }\right) \partial _{\omega }+%
\mathfrak{z}_{s}\omega _{s0}^{\prime 2}\partial _{\omega }^{2}\right] \left[
f(k,1-\omega ,\omega )\,\beta _{i}\left( k,1-\omega \right) \right]
\right\vert _{\omega =\omega _{s0}}].
\label{integral in rho d_k^2 rho-result}
\end{align}

Along similar lines we find the integral involving the third derivative in $%
k $: 
\begin{align}
I_{\rho \partial _{k}^{3}\rho }& =\int_{\allowbreak \eta }^{+\infty }\hspace{%
-2pt}dk\int_{-\infty }^{+\infty }\hspace{-2pt}d\omega f(k,\omega ,1-\omega
)\rho _{i}(\omega ,k)\partial _{k}^{3}\rho _{s}(1-\omega ,k)  \notag \\
& =\int_{\allowbreak \eta }^{+\infty }\hspace{-2pt}dk\left[ \left[ \mathfrak{%
z}_{s}^{\left( 3\right) }+\left( 3\mathfrak{z}_{s}^{\prime \prime }\omega
_{s0}^{\prime }+3\mathfrak{z}_{s}^{\prime }\omega _{s0}^{\prime \prime }+%
\mathfrak{z}_{s}\omega _{s0}^{\left( 3\right) }\right) \partial _{\omega
}\right. \right.  \notag \\
& +\left. \left. \left. \left( 3\mathfrak{z}_{s0}^{\prime }\omega
_{s0}^{\prime ^{2}}+3\mathfrak{z}_{s}\omega _{s0}^{\prime \prime }\omega
_{s0}^{\prime }\right) \partial _{\omega }^{2}+\mathfrak{z}_{s}\omega
_{s0}^{\prime ^{3}}\partial _{\omega }^{3}\right] \left[ f(k,1-\omega
,\omega )\,\beta _{i}\left( k,1-\omega \right) \right] \right\vert _{\omega
=\omega s0}\right] ,  \label{integral in rho d_k^3 rho-result}
\end{align}%
and\ the one involving the fourth derivative:%
\begin{align}
I_{\rho \partial _{k}^{4}\rho }& =\int_{\allowbreak \eta }^{+\infty }\hspace{%
-2pt}dk\int_{-\infty }^{+\infty }\hspace{-2pt}d\omega f(k,\omega ,1-\omega
)\rho _{i}(\omega ,k)\partial _{k}^{4}\rho _{s}(1-\omega ,k)  \notag \\
& =\int_{\allowbreak \eta }^{+\infty }\hspace{-2pt}dk\left[ \left[ \mathfrak{%
z}_{s}^{\left( 4\right) }+\left( 4\mathfrak{z}_{s}^{\left( 3\right) }\omega
_{s0}^{\prime }+6\mathfrak{z}_{s}^{\prime \prime }\omega _{s0}^{\prime
\prime }+4\mathfrak{z}_{s}^{\prime }\omega _{s0}^{\left( 3\right) }+%
\mathfrak{z}_{s}\omega _{s0}^{\left( 4\right) }\right) \partial _{\omega
}\right. \right.  \notag \\
& +\left( 6\mathfrak{z}_{s}^{\prime \prime }\omega _{s0}^{\prime ^{2}}+12%
\mathfrak{z}_{s}^{\prime }\omega _{s0}^{\prime \prime }\omega _{s0}^{\prime
}+\mathfrak{z}_{s}\left( 4\omega _{s0}^{\left( 3\right) }\omega
_{s0}^{\prime }+3\omega _{s0}^{\prime \prime ^{2}}\right) \right) \partial
_{\omega }^{2}+\left( 4\mathfrak{z}_{s}^{\prime }\omega _{s0}^{\prime ^{3}}+6%
\mathfrak{z}_{s}\omega _{s0}^{\prime \prime }\omega _{s0}^{\prime 2}\right)
\partial _{\omega }^{3}  \notag \\
& +\left. \left. \mathfrak{z}_{s}\omega _{s0}^{\prime ^{4}}\partial _{\omega
}^{4}\right] \left. \left[ f(k,1-\omega ,\omega )\,\beta _{i}\left(
k,1-\omega \right) \right] \right\vert _{\omega =\omega s0}\right] .
\label{integral in rho d_k^4 rho-result}
\end{align}

The next types of integrals we must handle are those coming from deriving
the spectral density $\rho _{s}(\omega _{s0}\left( p\right) -\omega ,k)$ in (%
\ref{scalar-damping-rate-longi}) and (\ref{scalar-damping-rate-trans}) with
respect to $p$. It is convenient to handle these terms by reintroducing $%
\delta \left( \omega _{s0}\left( p\right) -\omega -\omega ^{\prime }\right) $
and performing the derivative on it. The first such generic term is: 
\begin{align}
I_{\rho \rho \partial _{\omega }\delta }& =\int_{\allowbreak \eta }^{+\infty
}\hspace{-2pt}dk\int_{-\infty }^{+\infty }\hspace{-2pt}d\omega \int_{-\infty
}^{+\infty }\hspace{-2pt}d\omega ^{\prime }\,\partial _{\omega }\delta
\left( 1-\omega -\omega ^{\prime }\right) f(k,\omega ,\omega ^{\prime })\rho
_{i}(\omega ,k)\rho _{s}(\omega ^{\prime },k)  \notag \\
& =-\int_{\allowbreak \eta }^{+\infty }\hspace{-2pt}dk\int_{-\infty
}^{+\infty }\hspace{-2pt}d\omega \,\left. \partial _{\omega }\left[
f(k,\omega ,\omega ^{\prime })\rho _{i}(k,\omega )\right] \right\vert
_{\omega ^{\prime }=1-\omega }\rho _{s}(1-\omega ,k).
\label{integral in rho rho d_omega delta}
\end{align}%
Here too the residue-residue contribution is zero because of the same
kinematics. What remains to calculate is the cut-residue contribution, which
is equal to $-\int_{\allowbreak \eta }^{+\infty }\hspace{-2pt}%
dk\int_{-\infty }^{+\infty }\hspace{-2pt}d\omega \,\mathfrak{z}_{s}\partial
_{\omega }\left. \left[ f(k,\omega ,\omega ^{\prime })\beta _{i}\left(
k,\omega \right) \Theta \hspace{-1pt}\left( k-\left\vert \omega \right\vert
\right) \right] \right\vert _{\omega ^{\prime }=1-\omega }\delta \left(
1-\omega \mp \omega _{s0}\right) $ and straightforwardly shown to yield: 
\begin{equation}
I_{\rho \rho \partial _{\omega }\delta }=-\int_{\allowbreak \eta }^{+\infty }%
\hspace{-2pt}dk\,\mathfrak{z}_{s}\left. \partial _{\omega }\left[ f(k,\omega
,\omega _{s0})\beta _{i}(k,\omega )\right] \right\vert _{\omega =1-\omega
_{s0}}.  \label{integral in rho rho d_omega delta-result}
\end{equation}%
The other term involves a second order $\omega $-derivative; it is worked
out straightforwardly:%
\begin{eqnarray}
I_{\rho \rho \partial _{\omega }^{2}\delta } &=&\int_{\allowbreak \eta
}^{+\infty }\hspace{-2pt}dk\int_{-\infty }^{+\infty }\hspace{-2pt}d\omega
\int_{-\infty }^{+\infty }\hspace{-2pt}d\omega ^{\prime }\,\partial _{\omega
}^{2}\delta \left( 1-\omega -\omega ^{\prime }\right) f(k,\omega ,\omega
^{\prime })\rho _{i}(\omega ,k)\rho _{s}(\omega ^{\prime },k)  \notag \\
&=&-\int_{\allowbreak \eta }^{+\infty }\hspace{-2pt}dk\mathfrak{z}_{s}\left.
\partial _{\omega }^{2}\left[ f(k,\omega ,\omega _{s0})\beta _{i}(k,\omega )%
\right] \right\vert _{\omega =1-\omega _{s0}}.
\label{integral in rho rho d_omega^2 delta}
\end{eqnarray}%
Last are integrals involving an $\omega $-derivative over $\delta \left(
1-\omega -\omega ^{\prime }\right) $ and $k$-derivatives over $\rho _{s}$.
The treatment is similar and straightforward too. With one $k$-derivative we
obtain:%
\begin{eqnarray}
I_{\rho \partial _{k}\rho \partial _{\omega }\delta } &=&\int_{\allowbreak
\eta }^{+\infty }\hspace{-2pt}dk\int_{-\infty }^{+\infty }\hspace{-2pt}%
d\omega \int_{-\infty }^{+\infty }\hspace{-2pt}d\omega ^{\prime }\,\partial
_{\omega }\delta \left( 1-\omega -\omega ^{\prime }\right) f(k,\omega
,\omega ^{\prime })\rho _{i}(\omega ,k)\partial _{k}\rho _{s}(\omega
^{\prime },k)  \notag \\
&=&-\int_{\allowbreak \eta }^{+\infty }\hspace{-2pt}dk[\mathfrak{z}%
_{s}^{\prime }\left. \partial _{\omega }\left[ f(k,\omega ,\omega
_{s0})\beta _{i}(k,\omega )\right] \right\vert _{\omega =1-\omega _{s0}} 
\notag \\
&&-\mathfrak{z}_{s}\omega _{s0}^{\prime }\left. \partial _{\omega ^{\prime }}%
\left[ \partial _{\omega }\left[ f(k,\omega ,\omega _{s0})\beta
_{i}(k,\omega )\right] _{\omega =1-\omega ^{\prime }}\right] \right\vert
_{\omega ^{\prime }=\omega _{s0}}],
\label{integral in rho d_k rho d_omega delta}
\end{eqnarray}%
and with a second-order $k$-derivative we get:%
\begin{eqnarray}
I_{\rho \partial _{k}^{2}\rho \partial _{\omega }\delta }
&=&\int_{\allowbreak \eta }^{+\infty }\hspace{-2pt}dk\int_{-\infty
}^{+\infty }\hspace{-2pt}d\omega \int_{-\infty }^{+\infty }\hspace{-2pt}%
d\omega ^{\prime }\,\partial _{\omega }\delta \left( 1-\omega -\omega
^{\prime }\right) f(k,\omega ,\omega ^{\prime })\rho _{i}(\omega ,k)\partial
_{k}^{2}\rho _{s}(\omega ^{\prime },k)  \notag \\
&=&-\int_{\allowbreak \eta }^{+\infty }\hspace{-2pt}dk[\mathfrak{z}%
_{s}^{\prime \prime }\left. \partial _{\omega }\left[ f(k,\omega ,\omega
_{s0})\beta _{i}(k,\omega )\right] \right\vert _{\omega =1-\omega _{s0}} 
\notag \\
&&-2\mathfrak{z}_{s}^{\prime }\omega _{s0}^{\prime }\left. \partial _{\omega
^{\prime }}\left[ \partial _{\omega }\left[ f(k,\omega ,\omega _{s0})\beta
_{i}(k,\omega )\right] _{\omega =1-\omega ^{\prime }}\right] \right\vert
_{\omega ^{\prime }=\omega _{s0}}  \notag \\
&&-\mathfrak{z}_{s}\left( \omega _{s0}^{\prime \prime }\partial _{\omega
^{\prime }}+\omega _{s0}^{\prime 2}\partial _{\omega ^{\prime }}^{2}\right)
\left. \left[ \partial _{\omega }\left[ f(k,\omega ,\omega _{s0})\beta
_{i}(k,\omega )\right] _{\omega =1-\omega ^{\prime }}\right] \right\vert
_{\omega ^{\prime }=\omega _{s0}}].
\label{integral in rho d_k^2 rho d_omega delta}
\end{eqnarray}

These are generically all the types of integrals we will have to evaluate.
In each case, we have to examine the infrared behavior, i.e., the behavior
close to $\eta $. In the eventuality of the presence of an infrared
divergence, we have to extract it analytically. The remaining finite part of
the integral is generally performed numerically.

\subsubsection{Evaluation}

Let us work out a specific, general enough, example in some detail. Let us
take it from the coefficient of $p^{4}$ in the expression in (\ref%
{scalar-damping-rate-trans}) for $\gamma _{st}\left( p\right) $, namely the $%
\rho _{t}\left( \omega ,k\right) \partial _{k}\rho _{s}\left( 1-\omega
,k\right) $ contribution. The corresponding function $f$ is $f(k,\omega
,\omega ^{\prime })=k/\left( \omega \omega ^{\prime }\right) $. The integral
involved is of the type $I_{\rho \partial _{k}\rho }$, see (\ref{integral in
rho d_k rho-result}). Let us look at the first term $\mathfrak{z}%
_{s}^{\prime }f(k,1-\omega _{s0},\omega _{s0})\,\,\beta _{t}\left(
k,1-\omega _{s0}\right) $ where $\omega _{s0}$ stands here for $\omega
_{s0}\left( k\right) $. Using the expressions of $f$, the residue function $%
\mathfrak{z}_{s}\left( k\right) $ given after (\ref{rho_s}) and $\beta
_{t}\left( k,1-\omega _{s0}\right) $ given in (\ref{beta_l-t}), we perform a
small-$k$ expansion of the product to obtain the following small-$k$
behavior:%
\begin{equation}
\mathfrak{z}_{s}^{\prime }f(k,1-\omega _{s0},\omega _{s0})\beta _{t}\left(
k,1-\omega _{s0}\right) =\allowbreak -0.607927\,\allowbreak
/k+4.70205\,k+30.0885\,\allowbreak \allowbreak k^{3}\,+\allowbreak \mathcal{O%
}\left( k^{5}\right) .  \label{expansion-example}
\end{equation}%
It is clear that this is going to lead to a logarithmic divergence. To
extract this divergence from the integral, we split the latter into two
parts: one integration from $\eta $ to any finite number plus one second
integration from that finite number to $+\infty $. The integral writes then: 
\begin{align}
\int_{\eta }^{+\infty }dk\,\mathfrak{z}_{s}^{\prime }f(k,1-\omega
_{s0},\omega _{s0})\beta _{t}\left( k,1-\omega _{s0}\right) & =0.607927\ln
\eta -0.607927\,\ln \allowbreak k_{s0}(\ell )  \notag \\
& \hspace{-1.5in}+\int_{0}^{\allowbreak k_{s0}(\ell )}dk\left[ \mathfrak{z}%
_{s}^{\prime }f(k,1-\omega _{s0},\omega _{s0})\beta _{t}\left( k,1-\omega
_{s0}\right) +0.607927/k\right]  \notag \\
& \hspace{-1.5in}+\int_{\ell }^{1}dx\left. \mathfrak{z}_{s}^{\prime
}f(k,1-\omega _{s0},\omega _{s0})\beta _{t}\left( k,1-\omega _{s0}\right)
\right\vert _{k=\allowbreak k_{s0}(x)}.
\label{integral fzdkb in pl--1st expression}
\end{align}%
What we have done is this. In the finite part (the third term in (\ref%
{integral fzdkb in pl--1st expression})), we have changed the integration
variable from $k$ to $x\equiv k/\omega _{s0}(k)$, which implies that $%
k\equiv \allowbreak k_{s0}(x)$. Note that $k\rightarrow \infty $ implies $%
x\rightarrow 1$. The finite value in question that splits the original
integral into two parts is then chosen in terms of $x$ instead of $k$ and is
denoted by $\ell $. The second term in (\ref{integral fzdkb in pl--1st
expression}) is the original divergent piece of the total integral from the
integrand of which we have subtracted $-0.607927\,\allowbreak /k$ in order
to render it safe in the infrared, and hence the lower bound $\eta $ is
replaced by $0$. We must then of course add the contribution $%
-0.607927\,\int_{\eta }^{k_{s0}(\ell )}\frac{dk}{k}=0.607927\,\ln \eta
-0.607927\,\ln k_{s0}(\ell )$. As already mentioned, the finite value $\ell $
is arbitrary, but practically it must be chosen small enough in order to
make the integral $\int_{0}^{k_{s0}(\ell )}dk\left[ \mathfrak{z}_{s}^{\prime
}f(k,1-\omega _{s0},\omega _{s0})\,\,\beta _{t}\left( k,1-\omega
_{s0}\right) +0.607927/k\right] $ \textit{numerically} feasible. Indeed,
though we are assured of its finiteness \textit{analytically}, both
integrands $\mathfrak{z}_{s}^{\prime }f(k,1-\omega _{s0},\omega
_{s0})\,\,\beta _{t}\left( k,1-\omega _{s0}\right) $ and $0.607927/k$ are
still each (here logarithmically) divergent for small $k$. However, when $%
\ell $ is small, we can use a small-$k$ expansion in order to get a number
for the integral. We have pushed the expansion to $O(k^{11})$ and we start
having good convergence for already $\ell =0.6$. Also, we must (and do)
check that the final result does not depend on a particular value of $\ell $%
. We finally get: 
\begin{equation}
\int_{\eta }^{+\infty }dk\mathfrak{z}_{s}^{\prime }f(k,1-\omega _{s0},\omega
_{s0})\beta _{t}\left( k,1-\omega _{s0}\right) =0.607927\ln \eta +0.702549.
\label{integral fzdkb in pl--2nd expression}
\end{equation}

Now to the second contribution to the example we have chosen to detail,
which is $\int_{\eta }^{+\infty }dk\,\omega _{s0}^{\prime }\,\mathfrak{z}%
_{s}\left. \partial _{\omega }\left[ f(k,1-\omega ,\omega )\,\beta
_{i}\left( k,1-\omega \right) \right] \right\vert _{\omega =\omega _{s0}}$.
A small-$k$ expansion of the integrand yields: 
\begin{equation}
\omega _{s0}^{\prime }\,\mathfrak{z}_{s}\left. \partial _{\omega }\left[
f(k,1-\omega ,\omega )\,\beta _{t}\left( k,1-\omega \right) \right]
\right\vert _{\omega =\omega _{s0}}=-2.43171/k^{3}+24.8687/k-208.701k+%
\mathcal{O}(k^{3}).  \label{expansion w-prime-z-dw-fb in pl}
\end{equation}%
Here we have a $1/\eta ^{2}$ behavior in addition to the familiar $\ln \eta $%
. We therefore write: 
\begin{eqnarray}
\int_{\eta }^{+\infty }dk\,\omega _{s0}^{\prime }\,\mathfrak{z}_{s}\left.
\partial _{\omega }\left[ f(k,1-\omega ,\omega )\,\beta _{t}\left(
k,1-\omega \right) \right] \right\vert _{\omega =\omega _{s0}}
&=&-1.\,\allowbreak 215\,9/\eta ^{2}+1.\,\allowbreak 215\,9/k_{s}(\ell )^{2}
\notag \\
&&\hspace{-3.7in}-24.8687\ln \eta +24.8687\ln k_{s}(\ell
)+\int_{0}^{k_{s0}(\ell )}dk\left[ \omega _{s0}^{\prime }\,\mathfrak{z}%
_{s}\left. \partial _{\omega }\left[ f(k,1-\omega ,\omega )\,\beta
_{t}\left( k,1-\omega \right) \right] \right\vert _{\omega =\omega
_{s0}}\right.  \notag \\
&&\hspace{-3.7in}\left. +\frac{2.43171}{k^{3}}-\frac{24.8687}{k}\right]
+\int_{\ell }^{1}dx\,\omega _{s0}^{\prime }\,\mathfrak{z}_{s}\left. \partial
_{\omega }\left[ f(k,1-\omega ,\omega )\,\beta _{t}\left( k,1-\omega \right) %
\right] \right\vert _{k=k_{s0}(x)}.
\label{integral w-prime-z-dw-fb in pl-1st exppression}
\end{eqnarray}%
We have proceeded as already explained. The integrations are smooth and no
additional particular problem arises. Good convergence starts at $\ell =0.6$
and the $\ell $-independence of the sum is checked systematically. The final
result is: 
\begin{equation}
\int_{\eta }^{+\infty }dk\,\omega _{s0}^{\prime }\,\mathfrak{z}_{s}\left.
\partial _{\omega }\left[ f(k,1-\omega ,\omega )\,\beta _{t}\left(
k,1-\omega \right) \right] \right\vert _{\omega =\omega
_{s0}}=-1.\,\allowbreak 215\,9/\eta ^{2}-24.8687\ln \eta -27.2104.
\label{integral w-prime-z-dw-fb in pl-2nd exppression}
\end{equation}%
Putting these two contributions together, we arrive for this specific
example at:%
\begin{equation}
\left. I_{\rho \partial _{k}\rho }\right\vert _{\mathrm{example}%
}=-1.\,\allowbreak 215\,9/\eta ^{2}-24.26077\ln \eta -26.507851.
\label{example-final}
\end{equation}

All the terms are treated along similar steps: infrared divergences are
systematically detected by a small-$k$ expansion and extracted manually; the
finite-part integrals performed numerically. We sum all the contributions
and obtain exactly the two results (\ref{gamma-result}) for $\gamma
_{sl}\left( p\right) $ and $\gamma _{st}\left( p\right) $, namely:%
\begin{eqnarray}
\gamma _{sl}\left( p\right)  &=&\frac{e^{2}T}{16\pi }\left[
1.44253+0.309278\,\bar{p}^{2}-0.133949\bar{p}^{4}+\mathcal{O}\left( \bar{p}%
^{6}\right) \right] +\mathcal{O}\left( e^{3}T\right) ;  \notag \\
\gamma _{st}\left( p\right)  &=&\left. \hspace{-0.2cm}\frac{e^{2}T}{4\pi }%
\right[ -\left( 1.65937+1.62114\ln \allowbreak \bar{\eta}\right) \bar{p}^{2}
\notag \\
&&+\left. \left( 71.3264+57.4152\ln \allowbreak \bar{\eta}+\frac{1.94537}{%
\allowbreak \bar{\eta}^{2}}\right) \bar{p}^{4}+\mathcal{O}\left( \bar{p}%
^{6}\right) \right] +\mathcal{O}\left( e^{3}T\right) .
\label{gamma_sl+st-expansion}
\end{eqnarray}%
Note that by this method too the longitudinal contribution $\gamma
_{sl}\left( p\right) $ is free from any infrared divergence whereas the
transverse contribution is infrared sensitive.

\section{Scalar Energy}

Now we turn to the determination of the scalar energy $\omega _{s}\left(
p\right) $ to next-to-leading order in the coupling constant $e$. It is
defined in (\ref{next-to-leading-order-scalar-energy-1}). We want also to
obtain it in both the late and early $p$-expansion. We will see that in this
case too the same result is obtained, free from any infrared divergence.

\subsection{Late momentum expansion}

Recall that the next-to-leading order self-energy $^{\ast }\Sigma $ is the
sum of two diagrams $^{\ast }\Sigma _{1}$ and $^{\ast }\Sigma _{2}$, see (%
\ref{sigma_1+2}), (\ref{eff-self-energy-1}) and (\ref{eff-self-energy-2}).
Recall also that we have stated that $^{\ast }\Sigma _{2}$ is real. Indeed,
using the structure of the two-photon-two-scalar vertex (\ref{eff-ver-2})
and that of the photon propagator in the strict Coulomb gauge (\ref%
{effective-photon-propagator}), we have:%
\begin{equation}
^{\ast }\Sigma _{2}\left( P\right) =2e^{2}\,T\sum_{k_{0}}\int \frac{d^{3}k}{%
(2\pi )^{3}}\left[ \,^{\ast }\Delta _{l}(K)+2\,^{\ast }\Delta _{t}(K)\right]
.  \label{sigma-2}
\end{equation}%
The angular integrals over the $k$-solid angle and the sum over $k_{0}$ are
done as explained before. No $p$-expansion is needed. We obtain the
following expression:%
\begin{equation}
^{\ast }\Sigma _{2}\left( P\right) =\frac{\,e^{2}T}{\pi ^{2}}%
\int_{\allowbreak \eta }^{+\infty }dk\,k^{2}\int_{-\infty }^{+\infty }\frac{%
d\omega }{\omega }\left[ \rho _{l}(\omega ,k)\,+2\rho _{t}(\omega ,k)\right]
.  \label{real-part-sigma-2}
\end{equation}%
As just restated, real. On the contrary, $^{\ast }\Sigma _{1}$ has an
imaginary part which we have just manipulated in the previous section, both
with late and early $p$-expansions. One useful way to obtain the real part
of $^{\ast }\Sigma _{1}$ is to use its imaginary part and the following
dispersion relation \cite{kraemmer-rebhan-schulz}:%
\begin{equation}
\func{Re}\mathrm{Tr}\,^{\ast }\Delta _{l,t}\left( k_{0},k\right) \,^{\ast
}\Delta _{s}\left( q_{0},q\right) f\left( k\right) =\int\nolimits_{-\infty
}^{+\infty }\frac{dt}{t-\omega _{s0}}\frac{1}{\pi }\left[ \func{Im}\mathrm{Tr%
}\,^{\ast }\Delta _{l,t}\left( k_{0},k\right) \,^{\ast }\Delta _{s}\left(
q_{0},q\right) f\left( k\right) \right] _{\omega _{s0}=t}.
\label{dispersion-relation}
\end{equation}%
Using this dispersion relation and performing the Matsubara sum before any
momentum expansion, we obtain for the real part of the longitudinal
contribution $^{\ast }\Sigma _{1l}$ the following expression:%
\begin{equation}
\func{Re}\,^{\ast }\Sigma _{1l}(P)=-\frac{e^{2}T}{4\pi ^{2}}%
\int_{\allowbreak \eta }^{+\infty }\hspace{-4pt}\hspace{-6pt}%
dk\,k^{2}\int_{-\infty }^{+\infty }\hspace{-4pt}\frac{dt}{t-\omega _{s0}}%
\int_{-1}^{+1}\hspace{-6pt}dx\int_{-\infty }^{+\infty }\hspace{-4pt}d\omega 
\frac{\left( 2\omega _{s0}-\omega \right) ^{2}\,t\,}{\omega \left( t-\omega
\right) }\rho _{l}\left( \omega ,k\right) \rho _{s}\left( t-\omega ,q\right)
.  \label{real-part-sigma-l-1}
\end{equation}%
Remember, $x=\mathbf{\hat{k}.\hat{p}}$. For the transverse contribution $%
^{\ast }\Sigma _{1t}$, we obtain the following expression:%
\begin{eqnarray}
\func{Re}\,^{\ast }\Sigma _{1t}(P) &=&\frac{e^{2}T}{\pi ^{2}}%
p^{2}\int_{\allowbreak \eta }^{+\infty }dk\,k^{2}\int_{-\infty }^{+\infty }%
\frac{dt}{t-\omega _{s0}}\int_{-1}^{+1}dx\left( 1-x^{2}\right)   \notag \\
&&\times \int_{-\infty }^{+\infty }\frac{d\omega \,}{\omega \left( t-\omega
\right) }\,t\,\rho _{t}\left( \omega ,k\right) \rho _{s}\left( t-\omega
,q\right) .  \label{real-part-sigma-t-1}
\end{eqnarray}

The next step is to perform the integrals in (\ref{real-part-sigma-2}), (\ref%
{real-part-sigma-l-1}) and (\ref{real-part-sigma-t-1}). For $^{\ast }\Sigma
_{2}\left( P\right) $, we replace the photon dispersion relations by their
expressions in (\ref{rho_l-t}). The residue part leaves one integration over 
$k$ and the cut part restricts the integration over $\omega $ between $-k$
and $k$. Otherwise the integrations themselves are done numerically. No
early or late momentum expansion is necessary and no special problem arises.
In particular, the infrared behavior is safe. Recall that we have to
subtract the leading order terms for ultraviolet convergence as indicated
after (\ref{next-to-leading-dispersion}). From the longitudinal photon we
obtain $-0.183776\,e^{2}T$ and from the transverse photon $%
-0.000443967\,e^{2}T$, a much smaller contribution. Summing the two, we get:%
\begin{equation}
^{\ast }\Sigma _{2}\left( P\right) =-0.184\,22\,e^{2}T+\mathcal{O}\left(
e^{3}T\right) .  \label{sigma_2 final}
\end{equation}%
No $p$-dependence as announced previously, which means $^{\ast }\Sigma
_{2}\left( P\right) $ will only correct the scalar thermal mass $m_{s}$, a
correction free from any infrared divergence.

To perform the integrations in (\ref{real-part-sigma-l-1}) and (\ref%
{real-part-sigma-t-1}), we replace the scalar and photon spectral densities
by their respective expressions (\ref{rho_s}) and (\ref{rho_l-t}). Because
the scalar spectral density does not involve a cut, the integration over $t$
is trivial. Here too we could\ give values to the external momentum $p$ and
fit the resulting curves, but in the spirit of the present work and in view
of the coming comparison with the results from the early momentum expansion
method, we expand the integrand in powers of $p$ and perform analytically
the integration over $x$. The residue part in the photon spectral density
will only leave the integration over $k$ whereas the cut part will restrict
the integration over $\omega $. The remaining single and double integrals
are done numerically. Here too no special problem arises and the infrared
region is safe. After subtracting the corresponding leading-order terms for
ultraviolet convergence, we obtain:%
\begin{eqnarray}
\func{Re}\,^{\ast }\Sigma _{1l}(P) &=&e^{2}T\left[ -0.0746037-0.0201876\,%
\bar{p}^{2}+\mathcal{O}(\bar{p}^{4})\right] +\mathcal{O}\left( e^{3}T\right)
;  \notag \\
\func{Re}\,^{\ast }\Sigma _{1t}(P) &=&e^{2}T\left[ -0.165908\,\bar{p}^{2}+%
\mathcal{O}(\bar{p}^{4})\right] +\mathcal{O}\left( e^{3}T\right) .
\label{real sigma_1 final}
\end{eqnarray}%
It remains to sum (\ref{sigma_2 final}) and (\ref{real sigma_1 final}) and
divide by $2\omega _{s0}\left( p\right) $ as instructed in (\ref%
{next-to-leading-order-scalar-energy-1}) to obtain the scalar energy:%
\begin{equation}
\omega _{s}\left( p\right) =m_{s}\left[ \left( 1+0.258824\,e+\mathcal{O}%
\left( e^{2}\right) \right) +\left( 0.5+0.056684\,e+\mathcal{O}\left(
e^{2}\right) \right) \bar{p}^{2}+\mathcal{O}(\bar{p}^{4})\right] .
\label{scalar energy final}
\end{equation}

\subsection{Early momentum expansion}

Finally, let us recalculate the scalar energy $\omega _{s}\left( p\right) $
while introducing the expansion in $p$ at an early stage, before performing
the Matsubara sum. \ Since $^{\ast }\Sigma _{2}$ does not depend on $p$, it
will not be affected by this expansion; only $^{\ast }\Sigma _{1}$ is
affected. As in the calculation of $\gamma _{s}\left( p\right) $, the
integration over $x$ becomes straightforward. We do the Matsubara sum and
analytically continue to $\omega _{s0}\left( p\right) $. We obtain for the
longitudinal contribution: 
\begin{align}
\func{Re}\,^{\ast }\Sigma _{1l}(P)& =-\frac{e^{2}T}{2\pi ^{2}}%
\int_{\allowbreak \eta }^{+\infty }\hspace{-0.5cm}dk\,k^{2}\int_{-\infty
}^{+\infty }\frac{dt}{t-\omega _{s0}}\int_{-\infty }^{+\infty }\frac{d\omega 
}{\omega \left( t-\omega \right) }\,\,\rho _{l}(\omega ,k)  \notag \\
& \times t~\left( 2\omega _{s0}-\omega \right) ^{2}\left[ 1+\frac{p^{2}}{3}%
\left( \frac{1}{k}\partial _{k}+\frac{1}{2}\partial _{k}^{2}\right) +\dots %
\right] \rho _{s}(t-\omega ,k).  \label{real-part-sigma-l-2}
\end{align}%
Odd powers in $p$ cancel. The transverse contribution (\ref%
{real-part-sigma-t-1}) is already of order $p^{2}$ and does not need further 
$p$-expansion. The $x$-integral done, Matsubara sum and analytical
continuation to $\omega _{s0}\left( p\right) $ performed, we obtain:%
\begin{equation}
\func{Re}\,^{\ast }\Sigma _{1t}(P)=\frac{4e^{2}T}{3\pi ^{2}}%
p^{2}\int_{\allowbreak \eta }^{+\infty }dk\,k^{2}\int_{-\infty }^{+\infty }%
\frac{dt}{t-\omega _{s0}}\int_{-\infty }^{+\infty }\frac{d\omega }{\omega
\left( t-\omega \right) }\,t\,\rho _{t}(\omega ,k)\,\rho _{s}(t-\omega ,k).
\label{real-part-sigma-t-2}
\end{equation}

What remains now is to plug in the expressions of the spectral densities and
perform the integrals over $\omega $ and $t$, and then over the momentum $k$%
. The scalar spectral density $\rho _{s}$ has no cut part, which makes the
integration over $t$ simple. The residue part in $\rho _{l,t}$ eliminates
the integration over $\omega $ and the cut part restricts its limits.
Numerical work is needed for the rest. Here follows a display of generic
terms. First we have the type:%
\begin{eqnarray}
R_{\rho \rho } &=&\int_{\allowbreak \eta }^{+\infty }\hspace{-2pt}%
dk\int_{-\infty }^{+\infty }\hspace{-2pt}dt\int_{-\infty }^{+\infty }\hspace{%
-2pt}d\omega f(k,\omega ,t)\rho _{i}(k,\omega )\rho _{s}(k,t-\omega )  \notag
\\
&=&\int_{\allowbreak \eta }^{+\infty }dk\left[ \mathfrak{z}_{i}\,\mathfrak{z}%
_{s}\hspace{-2pt}\left[ f(k,\omega _{i},\omega _{i}+\omega
_{s0})\,\,-f(k,\omega _{i},\omega _{i}-\omega _{s0})-f(k,-\omega
_{i},-\omega _{i}+\omega _{s0})\right. \right.  \notag \\
&&\hspace{-42pt}+\left. f(k,-\omega _{i},-\omega _{i}-\omega _{s0})\right]
+\int_{-k}^{+k}d\omega \mathfrak{z}_{s}\beta _{i}\left( k,\omega \right) %
\left[ f(k,\omega ,\omega +\omega _{s0})\,\,-f(k,\omega ,\omega -\omega
_{s0})\right] ].  \label{R-rho-rho}
\end{eqnarray}%
The subscript $i$ stands for $l$ or $t$. Note that only integrals of type $%
I_{\rho \rho }$ contribute to the coefficient of zeroth order in $p$ in the
real part. The second type of integrals is the following: 
\begin{align}
R_{\rho \partial _{k}\rho }& =\int_{\allowbreak \eta }^{+\infty }\hspace{-2pt%
}dk\int_{-\infty }^{+\infty }\hspace{-2pt}dt\int_{-\infty }^{+\infty }%
\hspace{-2pt}d\omega f(k,\omega ,t)\,\rho _{i}(k,\omega )\,\partial _{k}\rho
_{s}(t-\omega ,k)  \notag \\
\hspace{-0.2cm}& =\int_{\allowbreak \eta }^{+\infty }dk\left[ \mathfrak{z}%
_{i}\,\mathfrak{z}_{s}^{\prime }\hspace{-2pt}\left[ f(k,\omega _{i},\omega
_{i}+\omega _{s0})\,\,-f(k,\omega _{i},\omega _{i}-\omega _{s0})-f(k,-\omega
_{i},-\omega _{i}+\omega _{s0})\right. \right.  \notag \\
& \hspace{-1cm}+\left. f(k,-\omega _{i},-\omega _{i}-\omega _{s0})\right]
+\omega _{s0}^{\prime }\,\mathfrak{z}_{i}\,\mathfrak{z}_{s}[\left. \partial
_{t}\left( f(k,\omega _{i},t)+f(k,-\omega _{i},-t)\right) \right\vert
_{t=\omega _{i}+\omega _{s0}}  \notag \\
& \hspace{-1cm}+\left. \partial _{t}\left( f(k,\omega _{i},t)+f(k,-\omega
_{i},t)\right) \right\vert _{t=\omega _{i}-\omega
_{s0}}]+\int_{-k}^{+k}d\omega \left[ \mathfrak{z}_{s}^{\prime }\beta
_{i}\left( k,\omega \right) \left[ f(k,\omega ,\omega +\omega
_{s0})\,\right. \right. \,  \notag \\
& \hspace{-1cm}-\left. f(k,\omega ,\omega -\omega _{s0})\right] +\omega
_{s0}^{\prime }\,\mathfrak{z}_{s}\beta _{i}\left( k,\omega \right) [\left.
\partial _{t}f(k,\omega ,t)\right\vert _{t=\omega +\omega _{s0}}+\left.
\partial _{t}f(k,\omega ,t)\right\vert _{t=\omega -\omega _{s0}}]]].
\label{R-rho-dk-rho}
\end{align}%
The third and last type of integrals is the one involving a second-order
derivative in $k$; it writes:%
\begin{align}
R_{\rho \partial _{k}^{2}\rho }& =\int_{\allowbreak \eta }^{+\infty }\hspace{%
-2pt}dk\int_{-\infty }^{+\infty }\hspace{-2pt}dt\int_{-\infty }^{+\infty }%
\hspace{-2pt}d\omega f(k,\omega ,t)\,\rho _{i}(k,\omega )\,\partial
_{k}^{2}\rho _{s}(t-\omega ,k)  \notag \\
\hspace{-0.2cm}& =\int_{\allowbreak \eta }^{+\infty }dk\left[ \mathfrak{z}%
_{i}\left( \hspace{-2pt}\mathfrak{z}_{s}^{\prime \prime }+\omega
_{s0}^{\prime 2}\,\mathfrak{z}_{s}\partial _{t}^{2}\right) \hspace{-2pt}%
\left[ \left. \left( f(k,\omega _{i},t)\right) \right\vert _{t=\omega
_{i}+\omega _{s0}}-\left. \left( f(k,\omega _{i},t)\right) \right\vert
_{t=\omega _{i}-\omega _{s0}}\right. \right.  \notag \\
& \hspace{-1cm}-\left. \left. \left( f(k,-\omega _{i},t)\right) \right\vert
_{t=-\omega _{i}+\omega _{s0}}+\left. \left( f(k,-\omega _{i},t)\right)
\right\vert _{t=-\omega _{i}-\omega _{s0}}\right] +\left( 2\omega
_{s0}^{\prime }\,\mathfrak{z}_{s}^{\prime }+\omega _{s0}^{\prime \prime }\,%
\mathfrak{z}_{s}\right) \mathfrak{z}_{i}  \notag \\
& \hspace{-1cm}\times \left[ \left. \partial _{t}\left( f(k,\omega
_{i},t)+f(k,-\omega _{i},-t)\right) \right\vert _{t=\omega _{i}+\omega
_{s0}}+\left. \partial _{t}\left( f(k,\omega _{i},t)+f(k,-\omega
_{i},t)\right) \right\vert _{t=\omega _{i}-\omega _{s0}}\right]  \notag \\
& \hspace{-1cm}+\int_{-k}^{+k}d\omega \left[ \left( \mathfrak{z}_{s}^{\prime
\prime }+\mathfrak{z}_{s}\omega _{s0}^{\prime 2}\partial _{t}^{2}\right)
\beta _{i}\left( k,\omega \right) \left[ \left. \partial _{t}^{2}f(k,\omega
,t)\right\vert _{t=\omega +\omega _{s0}}-\left. \partial _{t}^{2}f(k,\omega
,t)\right\vert _{t=\omega -\omega _{s0}}\right] \right.  \notag \\
& \hspace{-1cm}+\left. \left. \left( 2\omega _{s0}^{\prime }\,\mathfrak{z}%
_{s}^{\prime }+\omega _{s0}^{\prime \prime }\,\mathfrak{z}_{s}\right) \beta
_{i}\left( k,\omega \right) \left[ \left. \partial _{t}f(k,\omega
,t)\right\vert _{t=\omega +\omega _{s0}}+\left. \partial _{t}f(k,\omega
,t)\right\vert _{t=\omega -\omega _{s0}}\right] \right] \right] .
\label{R-rho-dk^2-rho}
\end{align}

Using these generic results, we perform the integrals numerically. We do not
encounter any additional problem and we find all the integrals safe in the
infrared. Here too we have to subtract the leading-order terms for
ultraviolet convergence and, putting all contributions together, the
numerical integrations yield the result:%
\begin{equation}
\omega _{s}\left( p\right) =m_{s}\left[ \left( 1+0.258824\,e+\mathcal{O}%
\left( e^{2}\right) \right) +\left( 0.5+0.056684\,e+\mathcal{O}\left(
e^{2}\right) \right) \bar{p}^{2}+\mathcal{O}(\bar{p}^{4})\right] .
\label{final result}
\end{equation}%
This is exactly the same result (\ref{scalar energy final}) obtained with
the late momentum expansion. We should note the fact that there is no
infrared sensitivity in the above expression, even though it has been
technically more involved to derive than the determination of the damping
rate $\gamma _{s}\left( p\right) $.

\section{Discussion}

This work aimed at calculating the damping rate $\gamma _{s}\left( p\right) $
and energy $\omega _{s}\left( p\right) $ for scalars in the context of
next-to-leading order hard-thermal-loop perturbation theory for scalar QED.
For each of the two quantities, we have carried out the calculation in two
ways: using both a late and early external momentum expansion, respectively
after and before the Matsubara sum is done and the analytic continuation to
real energies taken. Though technically different, both methods yield the
same results for the damping rate and energy. This fact is particularly
interesting in view of an early work \cite{AAB,AA} in which the early
momentum expansion was used in the context of next-to-leading order
hard-thermal-loop perturbation of hot QCD in order to extract analytically a
value for the non-moving longitudinal-gluon damping rate. The very
definition of that quantity imposed a systematic expansion in powers of the
external momentum $p$, at least to second order. From a technical point of
view, the question was where to perform the expansion. Ideally, the latest
possible, at least after the Matsubara sum and the analytic continuation to
real energies are done. But technically that was unfortunately not feasible
in QCD because of the complication of the intermediary steps, at least if
one wanted to carry through analytically. The early momentum expansion was
used instead, and legitimately the occurrence of infrared sensitivity of the
result was partially incriminated on the method used.

The present work shows that infrared divergences can occur without the early
momentum expansion. This is clear in the transverse contribution $\gamma
_{st}\left( p\right) $ to the scalar damping rate, see (\ref{gamma-result}).
Remember that up to order $p^{2}$, this result with the logarithmic
divergence is already found in \cite{thoma}. The reason we have pushed the
calculation to order $p^{4}$ is to demonstrate that other forms of
divergences do occur, here a $1/\eta ^{2}$. This was the case in QCD too.
Since these results are obtained without the early momentum expansion and
the same results are obtained with it, we think we have here a strong
indication that this latter is not responsible for any infrared sensitivity.

Furthermore, infrared sensitivity is not systematic when we perform a
momentum expansion, late or early. This is exemplified in the longitudinal
contribution $\gamma _{sl}\left( p\right) $ to the damping rate, see (\ref%
{gamma-result}), and, maybe more pertinently since the calculations are more
intricate, in the scalar energy $\omega _{s}\left( p\right) $. It is true
that only fourth order in $p$ is included in the determination of $\gamma
_{sl}\left( p\right) $ and second order in the determination of $\omega
_{s}\left( p\right) $, but the trend of the calculations indicates that
higher-order coefficients will eventually be safe.

A couple of objections may still rise. First, one could argue that in fact,
the late and early momentum expansions are actually the same for the
quantities we have treated since the expressions of these, before any
momentum expansion is performed, are still not analytically closed in terms
of $p$. To answer this objection, one can go to the photonic sector of
scalar QED treated in \cite{kraemmer-rebhan-schulz}. Indeed, the HTL-summed
next-to-leading order photon self-energies are\ evaluated in closed form for
all $\omega $ and $p$. Let us focus only on the longitudinal photons. Three
regions in $\omega $ and $p$ are to be distinguished: $\omega ^{2}<p^{2}$
(region I), $p^{2}<\omega ^{2}<4m_{s}^{2}+p^{2}$ (region II) and \ $%
4m_{s}^{2}+p^{2}<\omega ^{2}$ (region III). The longitudinal next-to-leading
order HTL-summed photon self-energy is found to have the following
expression \cite{kraemmer-rebhan-schulz}: 
\begin{equation}
\delta \,^{\ast }\Pi _{l}\left( \omega ,p\right) =\dfrac{e^{2}T}{8\pi }%
\dfrac{\omega ^{2}-p^{2}}{p^{2}}\left[ 4m_{s}+2i\varepsilon -i\dfrac{\omega
^{2}}{p}\ln \left( \dfrac{2m_{s}-i\left( \varepsilon +p\right) }{%
2m_{s}-i\left( \varepsilon -p\right) }\right) \right] ,
\label{HTL-summed photon self-energy}
\end{equation}%
where the prefix $\delta $ indicates here too that the known leading
hard-thermal-loop contribution has been subtracted for ultraviolet
convergence, and $\varepsilon =\Theta $ in regions I and III and $%
\varepsilon =i\left\vert \Theta \right\vert $ in region II, with $\Theta
^{2}\left( \omega ,p\right) =\omega ^{2}\left( \omega
^{2}-p^{2}-4m_{s}^{2}\right) /\left( \omega ^{2}-p^{2}\right) $. The full
longitudinal-photon dispersion relation is: 
\begin{equation}
\Omega _{l}^{2}=p^{2}+\Pi _{l}\left( \Omega _{l},p\right) ,
\label{full-dispersion-longitudinal}
\end{equation}%
where $\Pi _{l}$ is the full longitudinal-photon self-energy, and, up to
next-to-leading order, writes: 
\begin{equation}
\Omega _{l}^{2}\left( p\right) =\omega _{l}^{2}\left( p\right) +\dfrac{%
\delta \,^{\ast }\Pi _{l}\left( \omega _{l},p\right) }{1-\partial _{\omega
_{l}^{2}}\delta \Pi _{l}\left( \omega _{l},p\right) },
\label{next-to-leading-dispersion-longitudinal}
\end{equation}%
where $\delta \Pi _{l}\left( \omega ,p\right) $ is the hard thermal loop
given by \cite{kraemmer-rebhan-schulz}: 
\begin{equation}
\delta \Pi _{l}\left( \omega ,p\right) =3m_{p}^{2}\left( 1-\frac{\omega ^{2}%
}{p^{2}}\right) \left( 1-\frac{\omega }{2p}\ln \frac{\omega +p}{\omega -p}%
\right) ,  \label{photon-htl}
\end{equation}%
with $m_{p}=eT/3$ the photon thermal mass and $\omega _{l}\left( p\right) $
the on-shell longitudinal photon energy, solution to (\ref%
{full-dispersion-in-SQED}) to lowest order where only the hard thermal loop
is kept in the self-energy. Using (\ref{photon-htl}), we can rewrite (\ref%
{next-to-leading-dispersion}) as: 
\begin{equation}
\Omega _{l}^{2}\left( p\right) =p^{2}+\delta \Pi _{l}\left( \omega
_{l},p\right) +\dfrac{2\omega _{l}^{2}\left( p\right) }{3m_{p}^{2}+p^{2}-%
\omega _{l}^{2}\left( p\right) }\delta \,^{\ast }\Pi _{l}\left( \omega
_{l},p\right) .  \label{next-to-leading-dispersion-long-rewritten}
\end{equation}%
Remember that all these results involve no expansion in $p$ and the
Matsubara sum and analytic continuation to real energies are already done.
Also, all intermediary integrals are performed. Now we expand. The region of
interest to us is region II, where we are allowed to perform the following
expansion: 
\begin{equation}
\omega _{l}\left( p\right) =1+\frac{3}{10}\tilde{p}^{2}-\frac{3}{280}\tilde{p%
}^{4}+\mathcal{O}\left( \tilde{p}^{6}\right) ;\qquad \tilde{p}=p/m_{p}.
\label{expansion-omaga-SQED}
\end{equation}%
Using this, we perform the expansion of (\ref%
{next-to-leading-dispersion-rewritten}) in powers of $\tilde{p}$. It is
straightforward and we find: 
\begin{equation}
\Omega _{l}^{2}\left( p\right) =m_{p}^{2}\left[ \left( 1-0.368\,e+\mathcal{O}%
\left( e^{2}\right) \right) +\left( \frac{3}{5}-0.0536\,e+\mathcal{O}\left(
e^{2}\right) \right) \tilde{p}^{2}+\mathcal{O}(\tilde{p}^{4})\right] .
\label{Omega-expanded}
\end{equation}%
Here we have together the leading and next-to-leading orders in the coupling 
$e$. Note that there is no infrared sensitivity. What remains now is to
perform the expansion in powers of $\tilde{p}$ \textit{before} the Matsubara
sum and analytic continuation to real energies are done, and carry on with
steps similar to what we have gone through in this work. This we do and we
obtain exactly the \textit{same} result (\ref{Omega-expanded}). There is no
need to display the results we think. This is an additional indication that
the early momentum expansion works. Of course, when it can be avoided like
in the present context of scalar QED, there is no need to use it. But in the
context of hot QCD, a more realistic theory, it is sometimes necessary if
one wants to manipulate analytically.

Last, one may argue that if the calculations are to lead to infrared
sensitivity, then the regularization with an infrared cut-off $\eta $ should
not be used in the first place. To answer this, one can say that, first of
all, infrared sensitivity is not systematic in all similar quantities that
are treated in similar methods. We have seen a good example of this in this
work, which is the two contributions to the scalar damping rate $\gamma
_{s}\left( p\right) $: the longitudinal contribution $\gamma _{sl}\left(
p\right) $ is infrared safe and the transverse contribution $\gamma
_{st}\left( p\right) $ infrared sensitive, though both are calculated in
exactly the same manner. Hence infrared sensitivity may be a feature of the
physical quantity in question. Furthermore, we could have taken the
transverse contribution $\gamma _{st}\left( p\right) $, made no expansion in 
$p$ and tried to determine it numerically for different values of $p$
without introducing any infrared cutoff. Then the internal momentum $k$%
-integration would not converge, because precisely of the very infrared
sensitivity. One could argue that the regularization should be different
from the naive introduction of an infrared cut-off. But, in whatever way it
is chosen, regularization will only exhibit a potential infrared
sensitivity; it will not remove or cancel it. For that purpose, some other
procedure has to be invoked, and we have never meant to do that. This
particular issue would be pursued elsewhere.

\end{document}